\documentclass[12pt]{amsart}

\usepackage{amsmath}
\usepackage{amsfonts}
\usepackage{amssymb}
\usepackage{ifthen}

\theoremstyle{plain}
\newtheorem{Theorem}{Theorem}[section]
\newtheorem{Proposition}[Theorem]{Proposition}
\newtheorem{Lemma}[Theorem]{Lemma}
\newtheorem{Corollary}[Theorem]{Corollary}
\newtheorem{Remark}{Remark}
\newenvironment{Proof}[1][xxx]%
{{\ifthenelse{\equal{#1}{xxx}}{\scshape Proof.}{\scshape Proof #1.}}}{$\square$\par}

\newcommand{\NN}{\mathbb{N}}

\newcommand{\RR}{\mathbb{R}}
\newcommand{\CC}{\mathbb{C}}

\newcommand{\IM}{\operatorname{Im}}
\newcommand{\Kernel}{\operatorname{ker}}
\newcommand{\Image}{\operatorname{im}}
\newcommand{\Rank}{\operatorname{rank}}
\newcommand{\tr}{\operatorname{tr}}

\newcommand{\PDO}[1]{\ensuremath{\Psi^{#1}}}

\newcommand{\PDOt}[1]{\ensuremath{\Psi_t^{#1}}}
\newcommand{\Tstar}{\ensuremath{T^\star}}
\newcommand{\cTstar}{\ensuremath{\widetilde{T}^\star}}
\newcommand{\SP}[3]{\ensuremath{{\langle #2, #3\rangle}_{#1}}}
\newcommand{\SPn}[2]{\ensuremath{{|#2|}_{#1}}}

\newcommand{\WF}{\operatorname{WF}}
\newcommand{\WFs}[1]{\operatorname{WF}^{(#1)}}
\newcommand{\WFb}{\operatorname{WF}_b}

\newcommand{\WFpols}[1]{\operatorname{WF}^{(#1)}_{\operatorname{pol}}}
\newcommand{\Id}{\operatorname{Id}}

\newcommand{\Cinfty}{C^\infty}
\newcommand{\Dprime}{{\mathcal D}'}

\newcommand{\Hs}[1]{H^{(#1)}}
\newcommand{\Hcs}[1]{H_c^{(#1)}}
\newcommand{\boundary}{\partial}
\newcommand{\closure}[1]{\overline{#1}}
\newcommand{\interior}[1]{{#1}^{\circ}}
\newcommand{\gradient}{\nabla}
\newcommand{\divergence}{\nabla\cdot}
\newcommand{\restrictto}[1]{|_{#1}}
\newcommand{\Mod}[1]{\langle #1\rangle}
\newcommand{\Hyperbolic}{{\mathcal H}}
\newcommand{\Elliptic}{{\mathcal E}}
\newcommand{\Glancing}{{\mathcal G}}
\newcommand{\ScattRel}{{\mathcal S}}
\newcommand{\LameClass}{{\mathcal L}}

\newcommand{\In}{{\operatorname{in}}}
\newcommand{\Out}{{\operatorname{out}}}
\newcommand{\picked}[1]{{#1}^{(0)}}
\newcommand{\yzero}{y^{(0)}}
\newcommand{\etazero}{{\eta}^{(0)}}
\newcommand{\spacetime}{\RR\times\Omega}
\newcommand{\spacetimeclosure}{\RR\times\overline{\Omega}}
\newcommand{\spacetimebdry}{\RR\times\boundary\Omega}
\newcommand{\pdo}{pseudo-differential operator}
\newcommand{\Pdo}{Pseudo-differential operator}
\newcommand{\ropy}{\ensuremath{r_Y}} 

\begin{document}
\title[Elastodynamics with Residual Stress]{Propagation of Polarization in Elastodynamics with Residual Stress and Travel Times}

\author{S\"onke Hansen}
\address{\hskip-\parindent
S\"onke Hansen\\
Fachbereich Mathematik-Informatik\\ 
Universit\"at Paderborn\\ 
33095 Paderborn, Germany}

\author{Gunther Uhlmann}
\address{\hskip-\parindent
Gunther Uhlmann\\
Department of Mathematics\\ 
University of Washington\\ 
Seattle, WA 98195-4350, USA}

\thanks{The first author thanks the Department of Mathematics at the
   University of Washington for its hospitality during his visit in fall 2000. The second author is partly supported by NSF grant DMS-0070488 and a John
   Simon Guggenheim fellowship. The second author also thanks MSRI for partial
   support and for providing a very stimulating environment during the
   inverse problems program in fall 2001.}

\date{}

\maketitle

\section{Introduction}
\label{section:intro}

Consider an elastic medium which occupies a bounded domain
$\Omega\subset\RR^3$ with
smooth boundary $\boundary\Omega$ and exterior normal $\nu$.
Displacement is a time-dependent vector field
$u(t,\cdot)$ on $\closure{\Omega}$.
Small displacements satisfy, in a source-free medium,
the equations for (linearized) elastodynamics,
\begin{equation}
\label{elastodynamics}
\rho\, \partial^2 u/\partial t^2=\divergence S
\quad\text{with}\quad
S= R+\gradient u\, R+ C E.
\end{equation}
Here $0<\rho\in\Cinfty(\closure{\Omega})$ denotes the density, 
$S$ is the Piola-Kirchhoff stress tensor
which obeys the relation $S F^T =F S^T$ where
$F=I +\gradient u$ is the deformation gradient.
Divergence and transpose are taken with respect to
the Euclidean metric $|\cdot|$.
The elasticity tensor $C$ maps 
infinitesimal strain tensors $E=(\gradient u+{\gradient u}^T)/2$
to symmetric stress tensors $CE$.
$C$ represents the material
properties on the elastic medium.
$R(x)$, the residual stress tensor, is a symmetric $3\times 3$-matrix,
$\Cinfty$ on $\closure{\Omega}$.
It satisfies $\divergence R=0$.
See \cite[Sect.~23]{gurtin:84:linear-elasticity},
\cite{hoger:86:determ-residual}.
We call
\begin{equation}
\label{operator-of-linear-elastodynamics-with-residual-stress}
Pu= -\rho\, \partial^2 u/\partial t^2 +\divergence \big(R+\gradient u\, R+ C E\big)
\end{equation}
the operator for elastodynamics.
$P$ is isotropic if the elasticity tensor is as follows,
\begin{equation}
\label{isotropic-C}
CE = \lambda \tr(E) I +2\mu E
\quad\text{with $0<\lambda,\mu\in\Cinfty(\closure{\Omega})$.}
\end{equation}
$\lambda$ and $\mu$ are the Lam\'e parameters of the elastic medium.

The inverse problem for operators of elastodynamics
is to recover as much as possible of the elasticity tensor $C$
and of the residual stress tensor $R$ from measurements performed
at the space-time boundary $\spacetimebdry$.
See \cite{man/lu:87:acousto-residual} for the beginnings of an
acoustoelastic theory of residual stress determination based
on wave propagation methods.

We deal with an inverse problem for subclasses
$\LameClass(L,\varepsilon)$ of operators for isotropic elastodynamics.
Here $L,\varepsilon>0$, and, by definition,
$P \in\LameClass(L,\varepsilon)$ if and only if
\begin{equation}
\label{Lame-bound}
\lambda(x)+2 \mu(x), 1/\mu(x), 1/\rho(x)\leq L
\quad\text{when $x\in\closure{\Omega}$}
\end{equation}
and
\begin{equation}
\label{R-is-very-small}
|R(x)|\leq \varepsilon \mu(x)
\quad\text{when $x\in\closure{\Omega}$.}
\end{equation}
If $\varepsilon>0$ is sufficiently small the initial boundary
value problem is well-posed and microlocal parametrices exist.
Assumptions like~\eqref{R-is-very-small} with $\varepsilon$ small
have been introduced before to ascertain well-posedness.
See, e.g., \cite{robertson:97:boundary-ident-residual} for the static case.

The (hyperbolic) Dirichlet-to-Neumann map
\begin{equation}
\label{DN-map}
\Lambda:
u\restrictto{\spacetimebdry} \mapsto {\nu\cdot S}\restrictto{\spacetimebdry}
\end{equation}
encodes boundary measurements.
Here $u$ solves~\eqref{elastodynamics} with zero initial data.
We say that a property of an operator for elastodynamics from a given class is
\emph{determined by boundary measurements}
if the property is the same for any two operators
in the class with identical Dirichlet-to-Neumann maps.

A useful approach to inverse problems consists in
using high-frequency waves $u$ generated by boundary data
with singularities.
From travel times of singularities of $u$ 
recorded at $\boundary\Omega$
one then aims to recover the requested properties.
The latter problem is called an inverse kinematic problem.
Obviously, an important step is to prove that travel times are in fact
determined by boundary measurements.
The main goal of this paper is provide a result of this kind
which is applicable also when caustics may develop.

We study the propagation of polarization in the sense
of Dencker~\cite{dencker:82:polarization-sets}
for the initial boundary problem of 
the operator for isotropic elastodynamics, $P$.
In Proposition~\ref{P-is-rpt} we show that $P$ is a system
of real principal type if the residual stress $R$ satisfies
\begin{equation}
\label{R-is-small}
\mu(x){|\xi|}^2+ R(x)\xi\cdot\xi >0
\quad\text{when $(x,\xi)\in\Tstar(\closure{\Omega})\setminus 0$.}
\end{equation}
If \eqref{R-is-small} holds then
\begin{align}
\label{metric-gs}
\SP{S}{\xi}{\xi} &=\big(\mu(x){|\xi|}^2+ R(x)\xi\cdot\xi\big)/\rho(x),
\\
\label{metric-gp}
\SP{P}{\xi}{\xi} &=\big((\lambda(x)+2\mu(x)){|\xi|}^2+ R(x)\xi\cdot\xi\big)/\rho(x),
\end{align}
$(x,\xi)\in\Tstar(\closure{\Omega})$,
are the duals $g_{S/P}^{-1}$ of Riemannian metrics $g_{S/P}$ on $\closure{\Omega}$.
The characteristic variety of $P$ is the union of the
subvarieties $\tau^2-\SP{S}{\xi}{\xi}=0$ and $\tau^2-\SP{P}{\xi}{\xi}=0$
which correspond to shear and compressional waves, respectively.

The \emph{lens map} or scattering relation $\ScattRel$
of a metric $g$ on $\closure{\Omega}$ is defined as follows.
Consider bicharacteristic curves,
$\gamma:[a,b]\rightarrow\Tstar(\closure{\Omega}\times \RR)$,
of the Hamilton function
$H(t,x,\tau,\xi)=\tau^2-g^{-1}(x,\xi)$
which satisfy the following:
$\gamma(]a,b[)$ lies over the interior,
$\gamma$ intersects the boundary non-tangentially at
$\gamma(a)$ and $\gamma(b)$, and
time increases along $\gamma$.
By definition, $\ScattRel$ is the subset of 
$\big(\Tstar(\spacetimebdry)\setminus 0\big)^2$
obtained by projecting endpoint pairs $(\gamma(b),\gamma(a))$.
It is well-known that
$\ScattRel$ is a homogeneous canonical relation on $\Tstar(\spacetimebdry)\setminus 0$.
(See \cite{guillemin:76:sojourn} for the concept of a scattering relation.)
$\ScattRel$ is a diffeomorphism between open subsets of $\Tstar(\spacetimebdry)\setminus 0$.
We denote by $\ScattRel_S$ (resp.\ $\ScattRel_P$) the lens map of $g_S$ 
(resp.\ $g_P$) and call it the shear (resp.\ compressional) lens map.

Our main result is the following.
\begin{Theorem}
\label{thm:recover-relations}
Given $L>0$ there exists $\varepsilon>0$ such that
in the class $\LameClass(L,\varepsilon)$
the shear and the compressional lens maps are determined
by boundary measurements.
\end{Theorem}

Note that travel times of shear and compressional waves are recovered
separately from boundary measurements.

Let $g$ be Riemannian metric on $\closure{\Omega}$.
Denote by $D$ the open subset of
$\boundary\Omega\times\boundary\Omega$
which consists of the pairs $(x,y)$ of boundary points which
can be joined by a geodesic which passes through
the interior except for the endpoints $x$ and $y$ where
it intersects $\boundary\Omega$ transversally.
By definition, the boundary distance function of
$(\closure{\Omega},g)$ is the function $d:D\rightarrow[0,\infty[$
which assigns to $(x,y)\in D$ the geodesic distance, i.e.,
the infimum of the lengths of such geodesics.
If $(\closure{\Omega},g)$ is strictly convex then
$D$ is the complement of the diagonal and $d$ is smooth.
Geodesics of $g$ are projections of bicharacteristic
curves of $\tau^2-g^{-1}(x,\xi)=0$.
Geodesic distances equal travel times.
When $d$ is smooth it is a generating function of (a subset
of) the lens maps $\ScattRel$ of $g$, i.e.,
$((t_1,x_1,\tau,\xi_1), (t_0,x_0,\tau,\xi_0))\in \ScattRel$
if $t_1-t_0=d(x_1,x_0)$ and
$\xi_j=-\tau \partial d(x_1,x_0)/\partial x_j$ for $j=0,1$.
(See \cite{caratheodory:35:variationsrechnung},
\cite{guillemin/sternberg:77:geometrical-asymptotics}.)
Clearly, $\ScattRel$ determines $d$.
Hence we have the following corollary
of Theorem~\ref{thm:recover-relations}.
Here we call the boundary distance functions of
the metrics $g_S$ and $g_P$ the shear and the
compressional boundary distance functions $d_S$
and $d_P$, respectively.

\begin{Corollary}
\label{cor:recover-distances}
Given $L>0$ there exists $\varepsilon>0$ such that
in the class $\LameClass(L,\varepsilon)$
the shear and the compressional boundary distance functions
are determined by boundary measurements.
\end{Corollary}

Rachele~\cite{rachele:xx:uniq-aniso-elast} has a similar result
under additional assumptions which exclude conjugate points.
Note that our result allows the presence of conjugate points.

In the case $R=0$ the metrics $g_S$ and $g_P$ are
conformal to the Euclidean metric.
Mukhometov~\cite{mukhometov:82/sib:recon-metric}
solved the inverse kinematic problem for 
conformal classes of metrics under assumptions which
exclude conjugate points.
Corollary~\ref{cor:recover-distances} and, e.g.,
Croke's theorem \cite[Theorem~C]{croke:91:rigidity} imply
the following uniqueness result of Rachele.

\begin{Corollary}\emph{\cite[Theorem~1]{rachele:00/jde:ip-elast-uniq-interior}}
\label{Racheles-Theorem}
In the class of operators of isotropic elastodynamics
with vanishing residual stresses,
and with $(\closure{\Omega}, g_S)$,
$(\closure{\Omega}, g_P)$ strictly convex,
the compressional speeds and shear speeds,
$c_P=\sqrt{(\lambda+2\mu)/\rho}$ and
$c_S=\sqrt{\mu/\rho}$,
are determined by boundary measurements.
\end{Corollary}

If residual stresses do not vanish the metrics become anisotropic.
From Corollary~\ref{cor:recover-distances} and a
result of Stefanov-Uhlmann
on the anisotropic inverse kinematic problem
\cite[Theorem 1.1]{stefanov/uhlmann:98:rigidity}
we deduce the following result.

\begin{Corollary}
\label{thm:recover-metrics-using-Stefanov-Uhlmann}
There is a $C^{12}(\closure{\Omega})$ neighbourhood $U$
of the euclidean metric such that the following holds.
Let $P^{(1)}$ and $P^{(2)}$ be operators of isotropic elastodynamics.
Assume $\Lambda^{(1)}= \Lambda^{(2)}$.
Assume $\closure{\Omega}$ strictly convex with respect to the
metrics $g_S^{(j)}$ and $g_P^{(j)}$.
If $g_S^{(j)}, g_P^{(j)}\in U$ then
$g_S^{(1)} = \Psi_S^\star g_S^{(2)}$,
$g_P^{(1)} = \Psi_P^\star g_P^{(2)}$
with diffeomorphisms
$\Psi_S,\Psi_P:\closure{\Omega}\rightarrow\closure{\Omega}$ 
which leave the boundary fixed, i.e.,
$\Psi_S(x)=\Psi_P(x)=x$ if $x\in\boundary\Omega$.
\end{Corollary}

In \cite[Theorem 1.1]{stefanov/uhlmann:98:rigidity} an additional
flatness assumption at the boundary of $\Omega$ is made.
This assumption is superfluous in view of
\cite[Theorem 2.1]{lassas/sharafutdinov/uhlmann:xx:semiglobal}.

We prove Theorem~\ref{thm:recover-relations} in section~\ref{section:proof}.
The facts needed about propagation of singularities and polarizations
in non-glancing boundary problems for systems of real principal type
are proved in section~\ref{section:first} for first order systems.
These are applied to second order systems and to elastodynamics in
sections~\ref{section:second} and \ref{section:elast}, respectively.
In particular, section~\ref{section:elast} contains an analysis of the
Dirichlet-to-Neumann map $\Lambda$ and its pseudo-differential properties.

\section{Singularities of First Order Boundary Problems}
\label{section:first}

We summarize some facts from the microlocal theory of boundary problems.
The results are due to
Dencker~\cite{dencker:82:polarization-sets},
G\'erard~\cite{gerard:85:polarisation},
Melrose~\cite{melrose:81:transf-bdry}, and
Taylor~\cite{taylor:75:reflection}.

Let $Z$ an open subset of half-space $\closure{\RR_{+}}\times\RR^{n}$
equipped with coordinates $x\geq 0$ and $y=(y_1,\dots,y_n)$.
$\xi$ and $\eta=(\eta_1,\dots,\eta_n)$ are the dual coordinates
in cotangent space.
Denote the boundary and the interior of $Z$ by $Y$ and $\interior{Z}$, respectively.
$\Dprime(Z)$ denotes the space of extendible distributions on $\interior{Z}$.
\Pdo s of order at most $m$ on $Y$ and on $Z$ acting along $Y$ are written
$A(y,D_y)\in\PDO{m}(Y)$ and $B(x,y,D_y)\in\PDOt{m}(Z)$, respectively.
$S^m = S^m(Y\times\RR^n)$ and $S_t^m = S^m(Z\times\RR^n)$ are
the corresponding symbol spaces.
Elements of $\PDOt{m}(Z)$ are called tangential \pdo s.
Symbols are always assumed polyhomogeneous (classical).
\Pdo s will always be chosen properly supported.
We denote by $\ropy u=u\restrictto{Y}$ the restriction of $u\in\Dprime(Z)$,
when defined.

We consider $u\in{\Dprime(Z)}^K$ such that 
$P u\in{\Cinfty(Z)}^K$ where $P$ is a $K\times K$ system
of \pdo s which are differential with respect to $x$,
\begin{equation}
\label{system-P}
P=\sum_{j=0}^m P_j D_x^j
\quad\text{with}\quad
P_j= P_j(x,y,D_y)\in\PDOt{j}(Z),\quad
P_0=\Id_K.
\end{equation}
The boundary wavefront set $\WFb(u)$,
defined in \cite{melrose:81:transf-bdry},
is a closed subset
of the compressed cotangent bundle $\cTstar(Z)$.
If the $P_j$ are differential operators
then $Pu\in\Cinfty$ is a non-characteristic boundary problem and hence $u$
is normally regular in the sense of Melrose~\cite[II.9]{melrose:81:transf-bdry}.
Recall from \cite{melrose:81:transf-bdry} or
\cite[18.3]{hormander:85:the-analysis-3}
the following properties of a normally regular distribution $u$.
$u\in\Cinfty([0,\varepsilon[,\Dprime(\RR^n))$ locally near $Y$.
$Au$ is normally regular if $A$ is a tangential \pdo.
The boundary wavefront set
$\WFb(u)\subset \Tstar(Y)\cup\Tstar(\interior{Z})\subset\cTstar(Z)$.
$(y,\eta)\in\Tstar(Y)\setminus\WFb(u)$ if and only if
$Au\in\Cinfty(Z)$ for some operator 
$A=A(x,y,D_y)\in\PDOt{m}(Z)$ which is non-characteristic at $(0,y,\eta)$.
The polarization set $\WFpols{s}(u)$ is, by definition the intersection
of the sets
$$
{\mathcal N}_A = \big\{ (x,\xi;w)\in\Tstar(\interior{Z})\times \CC^K\;;\; \sigma(A)(x,\xi)w=0\big\}
$$
where $A\in{\PDO{0}}$ runs over all $1\times K$  systems such that $Au\in\Hs{s}(\interior{Z})$.
See \cite{dencker:82:polarization-sets} and \cite{gerard:85:polarisation}
for the precise definition and for results on the propagation of
polarization along Hamilton orbits.

Let $P$ as in \eqref{system-P} with principal symbol $p$.
Following Dencker~\cite[Definition~3.1]{dencker:82:polarization-sets}
we say that $P$ is of real principal type if, microlocally
near a given point, the characteristic variety is given by $q=0$
with a scalar symbol $q$ of real principal type and
if there exists a matrix-valued symbol, $\tilde{p}$
such that $\tilde{p}p=q\Id_K$.
If we assume $P$ of real principal type then
$H=H_q$ is a Hamilton field of the characteristic variety $V=q^{-1}(0)$ of $P$.
A point $(y,\eta)\in\Tstar Y\setminus 0$ is called glancing for $P$
if, with respect to the natural projection,
its preimage in $V\cap\Tstar_Y Z$ contains a point where $Hx = 0$,
else $(y,\eta)$ is called non-glancing for $P$.
Bicharacteristics intersect the boundary transversally
at non-glancing points.

We now specialize to first order systems, $m=1$.
Let $G=G(x,y,D_y)\in\PDOt{1}(Z)$ be an $N\times N$ matrix of
tangential \pdo s with homogeneous principal symbol $g$.
We assume that $D_x \Id_N - G$ is of real principal type.
We are interested in the singularities of
normally regular solutions of
\begin{equation}
\label{problem-first-order}
D_x w - G(x,y,D_y)w \equiv 0 \mod{{\Cinfty(Z)}^N}.
\end{equation}
Let $(\yzero,\etazero)\in\Tstar(Y)\setminus 0$ non-glancing
for $D_x \Id_N - G$.

The following decoupling lemma is due to
Taylor~\cite{taylor:75:reflection} in the case of simple real characteristics
and to G\'erard~\cite{gerard:85:polarisation} in the case of real principal type systems.

\begin{Lemma}
\label{decoupling}
In a conic neighbourhood $\Gamma$ of $(0,\yzero,\etazero)$,
the algebraic and geometric multiplicities of
the real eigenvalues of $g(x,y,\eta)$ are equal and constant.
There are homogeneous real-valued $\mu_1,\dots,\mu_J\in S_t^1$ which
enumerate, in $\Gamma$, the distinct real eigenvalues of $g$.
Let $N_j$ denote the multiplicity of $\mu_j$.
There is an elliptic $N\times N$ matrix $S\in\PDOt{0}$
such that microlocally near $(0,\yzero,\etazero)$,
\begin{equation}
\label{eq:decoupling:PS}
\big(D_x\Id_N-G\big) S \equiv S \big(D_x\Id_N-H\big) 
\quad\mod{\PDOt{-\infty}}.
\end{equation}
$H\in\PDOt{1}$ is a block matrix with non-zero entries
only on the diagonal,
\begin{equation}
\label{eq:decoupling:H}
H=
\left(\begin{array}{ccccc}
\mu_1(x,y,D_y)\Id_{N_1} &&&&\\
&\ddots &&&\\
&&\mu_J(x,y,D_y)\Id_{N_J} && \\
&&& E_{+} &\\
&&&& E_{-}
\end{array}\right).
\end{equation}
The imaginary parts of the eigenvalues of the principal symbols of
$E_+, E_-\in\PDOt{1}$ are positive and negative, respectively.
\end{Lemma}

\begin{Proof}
The following constructions hold in some conic neighbourhood
$\Gamma$ of $(0,\yzero,\etazero)$.
$\Gamma$ may become smaller as the proof proceeds.

Since $A=D_x\Id_N-G$ is of real principal type its characterictic variety
is $V=q^{-1}(0)$ with a scalar real principal type symbol $q$.
The non-glancing assumption implies $\partial q/\partial\xi\neq 0$
at points $(0,\yzero,\xi,\etazero)\in V$.
By the implicit function theorem,
the real eigenvalues of $g(x,y,\eta)$ are smooth homogeneous functions
$\mu_1(x,y,\eta)<\ldots<\mu_J(x,y,\eta)$ in $\Gamma$.
We extend them as homogeneous real valued symbols
$\mu_1,\ldots,\mu_J\in S_t^1(Z\times \RR^n)$.

Let $\mu$ be a real eigenvalue of
$\picked{g}=g(0,\yzero,\etazero)$.
We show that the geometric multiplicity of
$\mu$ equals its algebraic multiplicity,
\begin{equation}
\label{eq:decoupling:eigenspace}
\Kernel \big((\mu-\picked{g})^r\big)
  = \Kernel \big(\mu-\picked{g}\big),
\quad\forall r\in\NN.
\end{equation}

Let $a=\xi-g$ denote the principal symbol $A$.
By the non-glancing hypothesis $\partial/\partial\xi$
is transversal to the characteristic variety $\det a=0$
at $(\yzero,\etazero)$.
The intrinsic characterisation of real principal type
\cite[Prop.~3.2]{dencker:82:polarization-sets}
shows that $\partial a/\partial\xi=\Id$
maps the kernel of $a$ isomorphically
onto the cokernel of $a$ at $\xi=\mu$.
Hence
\begin{equation}
\label{eq:decoupling:kerneldisjointimage}
\Kernel (\mu-\picked{g}) \cap
   \Image (\mu-\picked{g}) = 0.
\end{equation}
Equation~\eqref{eq:decoupling:eigenspace} easily follows from
\eqref{eq:decoupling:kerneldisjointimage}.

Let $\gamma_1,\ldots,\gamma_J$ be non-intersecting closed positively oriented Jordan curves
in the complex plane such that $\gamma_j$ encloses $\mu_j(0,\yzero,\etazero)$
but no other eigenvalue of $\picked{g}$.
(To enclose means that the winding number is non-zero.)
\begin{equation}
\label{eq:decoupling:proj}
\pi_j(x,y,\eta) 
    = \int_{\gamma_j} {(\lambda-g(x,y,\eta)/|\eta|)}^{-1}\,\frac{d\lambda}{2\pi i}
\end{equation}
is the spectral projector onto the sum of generalized eigenspaces
associated with the eigenvalues enclosed by $\gamma_j$ of $g(x,y,\eta)/|\eta|$.
Clearly, $\Kernel \big(\mu_j-\picked{g}\big)\subset \Image\pi_j$.
By~\eqref{eq:decoupling:eigenspace} equality holds at $(0,\yzero,\etazero)$.
By~\cite[Prop.~3.2]{dencker:82:polarization-sets}
the dimension of $\Kernel \big(\mu_j-\picked{g}\big)$ is constant.
Also the rank of $\pi_j$ is constant in $\Gamma$.
Hence
\begin{equation}
\label{eq:decoupling:ker}
\Kernel (\mu_j-g) = \Image \pi_j
\quad\text{in $\Gamma$.}
\end{equation}

It follows from \eqref{eq:decoupling:ker} that
the geometric and the algebraic multiplicities of
the real eigenvalues of $g$ coincide everywhere in $\Gamma$.
Therefore we can find an elliptic $N\times N$ matrix
$s(x,y,\eta)\in S_t^{0}$ such that $s^{-1}gs\in S_t^1$
has, in $\Gamma$, the block structure of the principal symbol
of the operator $H$ claimed in~\eqref{eq:decoupling:H}.

Choose $S$ with principal symbol equal to $s$.
We obtain \eqref{eq:decoupling:PS} with the error class
$\PDOt{-\infty}$ replaced by $\PDOt{0}$, however.
We use the uncoupling technique of \cite{taylor:75:reflection}
to obtain $K\in\PDOt{-1}$ such that the error is 
$\PDOt{-\infty}$ if we replace $S$ by $S(\Id+K)$.
After doing this, however, $H$ will only satisfy a weaker form than \eqref{eq:decoupling:H}
with $\mu_j(x,y,D_y)\Id_{N_j}$ is replaced by
$\mu_j(x,y,D_y)\Id_{N_j}+M_j$ with some $M_j\in{\PDOt{0}}$.
By~\cite[Lemme 2.1.]{gerard:85:polarisation} there exist
elliptic $N_j\times N_j$ matrices $E_j\in\PDOt{0}$ such that
$$
\big((D_x-\mu_j(x,y,D_y))\Id_{N_j}-M_j\big) E_j
\equiv
E_j \big((D_x-\mu_j(x,y,D_y))\Id_{N_j}\big)
$$
holds modulo operators in $\PDO{-\infty}$.
Let $E\in\PDOt{0}$, $N\times N$, denote the diagonal block matrix
with blocks $E_1,\ldots,E_J$ and, in the lower right corner, $\Id$.
Finally, to remove the $M_j$'s, we replace $S$ by $SE$.
\end{Proof}

Let $B$ be a $K\times N$ matrix in $\PDOt{0}$ with homogeneous principal symbol $b$.
Given $h\in{\Dprime(Y)}^K$ we wish
to solve equation~\eqref{problem-first-order}
under the boundary condition specified by $B$ and $h$,
\begin{equation}
\label{bdry-problem-first-order}
\begin{aligned}
D_x w - Gw & \equiv 0 \mod{{\Cinfty(Z)}^N}, \\
Bw\restrictto{Y} & \equiv h\mod{{\Cinfty(Y)}^K}.
\end{aligned}
\end{equation}
Let $M_+\cup M_-=\{1,\dots,J\}$ be a disjoint union decomposing
the set of real eigenvalues of $g(0,\yzero,\etazero)$ into two parts.
We call the eigenvalue $\mu_j$ forward (resp.\ backward)
if $j\in M_+$ (resp.\ $j\in M_-$).
Correspondingly, we call characteristics and bicharacteristic
curves forward or backward.
In case $D_x\Id_N-G$ is hyperbolic with respect to a time
variable $t(x,y)$ such a decomposition arises as follows.
A bicharacteristic $\gamma$ issuing from the boundary into the interior
is forward (resp.\ backward) if $t$ increases (resp.\ decreases) along $\gamma$

We shall find a microlocal parametrix of the boundary
problem~\eqref{bdry-problem-first-order} if a
condition of Lopatinski type holds.
Define, for $(y,\eta)$ sufficiently close to $(\yzero,\etazero)$,
the forward Lopatinski space as the following linear
subspace of $\CC^N$,
\begin{equation}
\label{Lopatinski-forward}
L_g^+(y,\eta) = \Image\int_{\gamma^+}
     {(\lambda-g(0,y,\eta))}^{-1}\,d\lambda.
\end{equation}
$\gamma^+$ is a closed positively oriented Jordan curve
in the complex plane which encloses the eigenvalues of
$\picked{g}=g(0,\yzero,\etazero)$
which are real and forward or which have positive imaginary part.
$\gamma^+$ encloses no other eigenvalues of $\picked{g}$.

\begin{Proposition}
\label{parametrix-for-AB}
Assume that $b(0,\yzero,\etazero)$ maps
$L_g^+(\yzero,\etazero)$ onto $\CC^K$.
Then there exists a conic neighbourhood $\Gamma\subset\Tstar(Y)$
of $(\yzero,\etazero)$ and an operator
$W:{\Dprime(Y)}^K \rightarrow{\Dprime(Z)}^N$
such that the following holds.
For every $h\in {\Dprime(Y)}^K$ with $\WF(h)\subset \Gamma$
the distribution $w=Wh\in {\Dprime(Z)}^N$ is normally regular and
solves \eqref{bdry-problem-first-order}.
$\WF(w\restrictto{\interior{Z}})$ is contained in the union
of the forward bicharacteristics which issue from $\WF(h)$.
$W_0:=\ropy\circ W\in\PDO{0}(Y)$ is a $N\times K$ \pdo.
The principal symbol $w_0(y,\eta)$ of $W_0$ maps $\CC^K$ into
$L_g^+(y,\eta)$ and satisfies $b w_0=\Id$ in $\Gamma$.
\end{Proposition}

\begin{Proof}
Let $S$ and $H$ as in Lemma~\ref{decoupling}, and
denote their principal symbols by $s$ and $h$, respectively.
Clearly, $gs=sh$.
Hence $L_g^+=s L_h^+$.
The block structure of $H$ and the partitioning into forward
and backward eigenvalues defines a projector $\Pi$ on $\CC^N$.
$\Pi$ projects onto the subspace corresponding to the
blocks $\mu_j(x,y,D_y)\Id_{N_j}$, $j\in M_+$, and $E_+$ of $H$ along
the subspace corresponding to the
blocks $\mu_j(x,y,D_y)\Id_{N_j}$, $j\in M_-$, and $E_-$ of $H$.
Notice that $L_h^+=\Pi\CC^N$.
By assumption
\begin{equation}
\label{bsPi-surjective}
\CC^K= b L_g^+ = b s L_h^+ = b s \Pi \CC^N.
\end{equation}

The Cauchy problems
\begin{equation*}
(D_x -\mu_j(x,y,D_y))v \in \Cinfty(Z), \quad
v\restrictto{Y} \in\Cinfty(Y),
\end{equation*}
are solved using scalar Fourier integral operators
$V_j$, \cite{duistermaat:73:fio}.
The wavefront set of the solution $v=V_j f$ is contained
in the image of the bicharacteristics associated with $\xi-\mu_j(x,y,\eta)=0$
which issue from $\WF(f)$.
The parabolic system
\begin{equation*}
D_x v -E_{+}v \in{\Cinfty(Z)},\quad
v\restrictto{Y} \in{\Cinfty(Y)},
\end{equation*}
is solved using a Poisson operator $V_+$,
\cite{taylor:75:reflection}.
The solution $v=V_+ f$ has no singularities in $\interior{Z}$.
Therefore we may construct an operator
$V:{\Dprime(Y)}^N \rightarrow{\Dprime(Z)}^N$
such that the following holds for any $f\in {\Dprime(Y)}^N$.
$v=Vf$ is normally regular, $D_x v- Hv\in{\Cinfty(Z)}^N$,
and $\WF(v\restrictto{\interior{Z}})$ is contained in the union
of the forward bicharacteristics which issue from $\WF(f)$.
Furthermore, modulo ${\Cinfty(Y)}^N$, $v\restrictto{Y}\equiv \Pi f$.

$\ropy BSV$ is a $K\times N$ system of \pdo s on $Y$.
Its principal symbol $bs\Pi$ is, close to $(\yzero,\etazero)$,
surjective by \eqref{bsPi-surjective}.
Choose a $N\times K$ operator $C\in\PDO{0}(Y)$ which is a
right inverse, $\ropy BSVC\equiv B\ropy SVC\equiv \Id$.
$W=SVC$ satisfies the claims.
\end{Proof}

\begin{Remark}
If the boundary data $f$ is a Lagrangian distribution
then the solution $w=Wf$ is Lagrangian with respect to the
forward characteristics.
R\"ohrig \cite{roehrig:xx:dissertation} derives the transport equations
for the principal symbol of $w$ along the bicharacteristics.
\end{Remark}

To prepare waves with specified polarization we need
the following result about propagation of polarization
at the boundary.
Essentially this is a corollary of
\cite[Th\'eor\`eme 6.1]{gerard:85:polarisation}.

\begin{Proposition}
\label{WF-in-boundary-trace}
Let $w\in{\Dprime(Z)}^N$ normally regular such that
$(\yzero,\etazero)\notin\WFb(D_x w-Gw)$.
Assume $w\restrictto{Y}\in \Hs{s-1}(Y)$, $s>1$.
Let $\mu\in S^1$ be a real eigenvalue of $g(0,\cdot)$
in a conic neighbourhood of $(\yzero,\etazero)$.
Let $Q\in\PDO{0}(Y)$ with principal symbol equal to, in a neighbourhood
of $(\yzero,\etazero)$, the spectral projector
on the eigenspace of the eigenvalue $\mu$.
Then $(\yzero,\etazero)\in \WFs{s}(Qw\restrictto{Y})$ if and only if
$\WFpols{s}(w)$ contains a Hamilton orbit above the $\mu$-bicharacteristic
which issues from $(\yzero,\etazero)$.
\end{Proposition}

\begin{Proof}
Choose a parametrix $S^{-1}$ of $S$ in Lemma~\ref{decoupling}
and put $w'=S^{-1} w$.
The hypotheses of the Proposition
still hold with $w$ replaced by $w'$ and with $G$ replaced
by $H$ of~\eqref{eq:decoupling:H}.
Let $Q_\mu$ denote the projection to the components of the
block which corresponds to $\mu$ in the block decomposition
\eqref{eq:decoupling:H}.
Then $Q-Q_\mu\in\PDO{-1}(Y)$ and, using the assumption on $w\restrictto{Y}$,
$\WFs{s}(Q_\mu w\restrictto{Y}) = \WFs{s}(Qw\restrictto{Y})$.
$v=Q_\mu w$ solves the diagonal system
$(\yzero,\etazero)\notin\WFb(D_x v-\mu(x,y,D_y)v)$.
The assertion follows from well-known results on propagation of singularities
in the Cauchy problem for scalar strictly hyperbolic equations
and from~\cite[Theorem~4.2]{dencker:82:polarization-sets}.
\end{Proof}

\section{Second Order Boundary Problems}
\label{section:second}

Here we reduce the Dirichlet problem for second order
real principal systems to a boundary problem for
a first order real principal type system.

Let $P= D_x^2\Id_K+P_1(x,y,D_y) D_x+P_2(x,y,D_y)$ be
a $K\times K$ matrix of differential operators of second order.
We are interested in the Dirichlet problem
\begin{equation}
\label{bdry-problem-second-order}
\begin{aligned}
Pu & \equiv 0 \mod{\Cinfty(Z)}^K, \\
u\restrictto{Y} &\equiv f \mod{\Cinfty(Y)}^K.
\end{aligned}
\end{equation}
Any solution $u$ is normally regular.

We associate with \eqref{bdry-problem-second-order} an equivalent
first order boundary problem \eqref{bdry-problem-first-order} as follows.
Set $N=2K$,
$G\in\PDOt{1}$ the $N\times N$ matrix
\begin{equation}
\label{second-to-first-order}
G=\left(\begin{array}{cc}
0 & \Mod{D_y}\Id_K \\
- P_2 \,{\Mod{D_y}}^{-1} & -P_1 \\
\end{array}\right),
\end{equation}
and $B\in\PDOt{0}$ the $K\times N$ matrix with
$Bw=w_1$, $w=(w_1,w_2)$.
Here $\Mod{D_y}\in\PDOt{1}$ denotes the operator
with full symbol $\Mod{\eta}={(1+{|\eta|}^2)}^{1/2}\in S_t^{1}$.

\begin{Lemma}
\label{equivalence-of-bdry-problems}
Let $f\in{\Dprime(Y)}^K$ and $h=\Mod{D_y}f$.
Solutions $u$ of \eqref{bdry-problem-second-order}
and $w=(w_1,w_2)$ of \eqref{bdry-problem-first-order} are
related as follows.
If $u$ solves \eqref{bdry-problem-second-order}
then $w=(w_1,w_2)=(\Mod{D_y}u,D_x u)$
solves \eqref{bdry-problem-first-order}.
Conversely, if $w$ solves \eqref{bdry-problem-first-order}
then $u=\Mod{D_y}^{-1} w_1$
solves \eqref{bdry-problem-second-order}.
\end{Lemma}

\begin{Proof}
The first statement follows immediately from the definition
of $G$ and $B$.
For the proof of the converse statement let $w$ be a solution 
of \eqref{bdry-problem-first-order}.
The first row of $D_x w\equiv G w$ and the ellipticity
of $\Mod{D_y}$ imply $D_x u\equiv w_2$.
Hence the second row implies $Pu\equiv 0$.
By our choice of $B$ and $h$ the boundary conditions are
equivalent: $\Mod{D_y}u = Bw\equiv h =\Mod{D_y}f$.
\end{Proof}

Let $(x,y,\xi,\eta)\in\Tstar(Z)$, $\eta\neq 0$.
Let $p=\xi^2+p_1 \xi+p_2$ denote the principal symbol of $P$.
Then the principal symbol of $G$ is
$$
g= \left(\begin{array}{cc}
0 & |\eta| \\ -p_2/{|\eta|} & -p_1
\end{array}\right).
$$

\begin{Lemma}
\label{reduction-to-first-order}
Let $\eta\neq 0$. Then
\begin{equation}
\label{g-related-to-p}
(\xi-g')(\xi-g) = \left(\begin{array}{cc}p& \\ &p\end{array}\right)
\quad\text{where}\quad
g'= \left(\begin{array}{cc}
-p_1 & -|\eta| \\ p_2/{|\eta|} & 0
\end{array}\right)
\end{equation}
and
\begin{equation}
\label{map-kernels}
\ker(\xi-g) =
\left( \begin{array}{c}
|\eta|\Id_K \\ \xi \Id_K
\end{array}\right)
\ker(p).
\end{equation}
The characteristic varieties of $P$ and $D_x\Id_N-G$ are equal.
If $P$ is of real principal type then so is $D_x\Id_N-G$.
\end{Lemma}

\begin{Proof}
Equation~\eqref{g-related-to-p} is verified by direct computation.
Clearly, $(\xi-g)w=0$ with $w=(w_1,w_2)$, holds if and
only if $\xi w_1=|\eta|w_2$ and $p w_2=0$.
To prove the last assertion assume there is a $K\times K$ matrix
of symbols, $\tilde{p}$, such that $\tilde{p}p=q\Id_K$ holds
with a scalar real principal type smbol $q$.
Then, using \eqref{g-related-to-p}, we obtain a $N\times N$ matrix of
symbols, $\tilde{a}$, such that $\tilde{a}(\xi-g)=q\Id_N$.
\end{Proof}

\begin{Remark}
\label{Remark-reduction-to-first-order}
Assume $P$ of real principal type.
Let $C_0 u= (\Mod{D_y} u\restrictto{Y}, D_x u\restrictto{Y})$
denote the Cauchy data of a solution of $Pu\equiv 0$.
It follows from Proposition~\ref{WF-in-boundary-trace}
and Lemma~\ref{reduction-to-first-order} that
$\WFpols{s+1}(u)$ contains a Hamilton orbit above a given bicharacteristic
issuing from $\gamma=(y,\eta)\in\Tstar(Y)\setminus 0$
if and only if $\gamma\in\WFs{s}(QC_0 u)$ where $Q\in\PDO{0}(Y)$
with principal symbol equal to the spectral projector onto the
eigenspace $\{(\Mod{\eta}a,\xi a);\,p(0,y,\xi,\eta)a=0\}$
which corresponds to the given characteristic.
\end{Remark}

We give sufficient conditions for the existence of
a microlocal parametrix for the boundary problem
\eqref{bdry-problem-second-order}.

\begin{Proposition}
\label{singularities-in-Dirichlet}
Assume $P$ of real principal type.
Let $(\yzero,\etazero)\in\Tstar Y\setminus 0$ be non-glancing for $P$.
Let $\gamma^+$ be a closed positively oriented Jordan curve which
does not meet the poles of
$\lambda\mapsto {p(0,\yzero,\lambda,\etazero)}^{-1}$
and which has winding number $1$ (resp.\ $0$) with respect to
the poles with positive (resp.\ negative) imaginary part.
Assume that
\begin{align}
\label{Lopatinski-for-Dirichlet-g}
K & \geq \Rank\int_{\gamma^+} {\big(\lambda-g(0,\yzero,\etazero)\big)}^{-1}\,d\lambda, \\
\label{Lopatinski-for-Dirichlet-p}
K & \leq \Rank\int_{\gamma^+} {p(0,\yzero,\lambda,\etazero)}^{-1}\,d\lambda.
\end{align}
Then there exists a conic neighbourhood $\Gamma\subset\Tstar(Y)$
of $(\yzero,\etazero)$ and an operator
$U:{\Dprime(Y)}^K \rightarrow{\Dprime(Z)}^K$
such that for any $f\in {\Dprime(Y)}^K$ with $\WF(f)\subset \Gamma$
the distribution $u=Uf\in {\Dprime(Z)}^K$ is normally regular and
solves \eqref{bdry-problem-second-order}.
$\WF(u\restrictto{\interior{Z}})$ is contained in the union
of the forward bicharacteristics which issue from $\WF(f)$.
$U':=\ropy D_x U\in\PDO{1}(Y)$ is a $K\times K$ \pdo\ with
principal symbol $u'$ which satisfies
\begin{equation}
\label{symbol-of-DxU}
u'(\yzero,\etazero) \,
    \int_{\gamma^+} {p(0,\yzero,\lambda,\etazero)}^{-1}\,d\lambda
  = \int_{\gamma^+} \lambda\,{p(0,\yzero,\lambda,\etazero)}^{-1}\,d\lambda.
\end{equation}
\end{Proposition}

\begin{Proof}
We use the equivalence,
stated in Lemma~\ref{equivalence-of-bdry-problems},
of \eqref{bdry-problem-second-order}
with the first order boundary problem \eqref{bdry-problem-first-order}.

A real eigenvalue of $g(0,\yzero,\etazero)$ is, by definition,
forward if it is enclosed by $\gamma^+$.
First we show that our assumptions imply the following formula
for the Lopatinski space,
\begin{equation}
\label{Lg-expressed-with-p}
L_g^+(\yzero,\etazero) =
   \Image
   \left(\begin{array}{c} 
     \int_{\gamma^+} |\etazero|\, {p(0,\yzero,\lambda,\etazero)}^{-1}\,d\lambda \\
     \int_{\gamma^+} \lambda\, {p(0,\yzero,\lambda,\etazero)}^{-1}\,d\lambda
   \end{array}\right).
\end{equation}
From \eqref{g-related-to-p} we infer that the resolvent of $g$ is 
\begin{equation}
\label{symbol-of-resolvent}
{(\lambda-g)}^{-1}=
\left(\begin{array}{cc} * & |\eta| {p(\lambda)}^{-1} \\
     * & \lambda {p(\lambda)}^{-1}
 \end{array}\right)
\quad\text{where $p(\lambda)=\lambda^2+p_1 \lambda+p_2$, $\lambda\in\CC$.}
\end{equation}
Hence the right hand side in \eqref{Lg-expressed-with-p}
is contained in the left hand side.
Equality follows from the dimension
assumptions~\eqref{Lopatinski-for-Dirichlet-g}
and~\eqref{Lopatinski-for-Dirichlet-p}.

The principal symbol of $B$ is $b=(\Id_K,0)$.
Therefore \eqref{Lg-expressed-with-p},
\eqref{Lopatinski-for-Dirichlet-g}, and
\eqref{Lopatinski-for-Dirichlet-p} imply
$b L_g^+ = \CC^K$ at $(0,\yzero,\etazero)$.
Proposition~\ref{parametrix-for-AB} applies to give
a solution operator of \eqref{bdry-problem-first-order},
$W=(W_1,W_2)$ with $W_1=BW$.
Define $U= {\Mod{D_y}}^{-1} W_1 \Mod{D_y}$.
It follows from Lemma~\ref{equivalence-of-bdry-problems}
that $PU\equiv 0$4s, and $\ropy U\equiv \Id$, and
$W_2\Mod{D_y}\equiv D_x U$.
Hence $(\Mod{D_y},U')\equiv \ropy W\Mod{D_y} \in\PDO{1}(Y)$.
The principal symbol of $\ropy W$,
$(\Id_K, {|\eta|}^{-1} u'(y,\eta))$,
maps $\CC^K$ into the Lopatinski space $L_g^+(y,\eta)$.
Now we can read the formula~\eqref{symbol-of-DxU} 
off the equation~\eqref{Lg-expressed-with-p}.

The bound on $\WF(u\restrictto{\interior{Z}})$
follows from the bound on 
$\WF(w\restrictto{\interior{Z}})$
in Proposition~\ref{parametrix-for-AB}.
\end{Proof}

\section{Isotropic Elastodynamic Equations}
\label{section:elast}

In following $P$ denotes an operator for isotropic
elastodynamics introduced in
\eqref{operator-of-linear-elastodynamics-with-residual-stress}
and \eqref{isotropic-C} such that \eqref{R-is-small} holds.

The boundary problem $Pu=g$ in $\spacetime$, and
$u=0$ on $\spacetimebdry$, has the variational formulation:
$(\rho\ddot{u},v)+a(u,v)+(g,v)=0$,
$\forall v\in \big(\overset{\circ}{\Hs{1}}(\Omega)\big)^3$.
Here $a=a_0+a_R$,
$a_0(u,v)=\int_\Omega \tr \big(CE(u)\, {E(v)}^T\big)\,dx$, and
$a_R(u,v)=\int_\Omega \tr \big((\nabla u)\, R\, (\nabla v)^T\big) \,dx$.
It follows from Korn's inequality that
$a_0$ satisfies a coerciveness estimate $|a_0(u,u)|\geq c\|u\|_1^2$
with a positive constant depending
only on $\Omega$ and on a lower bound on the Lam\'e coefficient $\mu$,
\cite{duvaut/lions:76:inequalities}.
Given $L>0$ there exists $\varepsilon>0$ such that
$a$ is coercive if $P\in\LameClass(L,\varepsilon)$.
In fact, $a_R$ is absorbed into the coerciveness estimate if
\eqref{R-is-very-small} is assumed with $0<\varepsilon=\varepsilon(L)$
sufficiently small.
We use \cite[Thm.~III.4.1.]{duvaut/lions:76:inequalities}
to conclude that the initial boundary value problem for $P$ with
Dirichlet boundary conditions in ${\Hcs{s}(\spacetimebdry)}^3$, $s\geq 3$, is well-posed.
In particular, the Dirichlet-to-Neumann (DN) map~\eqref{DN-map} is defined,
\begin{equation}
\label{DN-mapping-property}
\Lambda: \Hcs{s+1}(\spacetimebdry)^3
   \rightarrow \Hs{s}(\spacetimebdry)^3
\quad\text{if $s\geq 2$.}
\end{equation}

Let $(t,x,\tau,\xi)$ denote a generic point
in $\Tstar(\spacetimeclosure)\setminus 0$.
The Euclidean metric $\xi^2=\xi\cdot\xi$ is used
to identify tangent and cotangent vectors of $\closure{\Omega}$.
For $\eta, \zeta\in\CC^3$ the dot product is the
analytic (non-Hermitian) extension,
$\eta\cdot\zeta= \eta_1\zeta_1 + \eta_2\zeta_2 + \eta_3\zeta_3$.
$\pi=\pi(\xi)=(\xi\otimes\xi)/(\xi\cdot\xi)$ denotes
the orthogonal projection onto a nonzero direction if $\xi\in\RR^3$.

As a consequence of \eqref{R-is-small} the metrics
defined in \eqref{metric-gs} and \eqref{metric-gp} satisfy
\begin{equation}
\label{gP-less-gS}
0<\SP{S}{\xi}{\xi} <\SP{P}{\xi}{\xi}
\quad\text{if $\xi\neq 0$.}
\end{equation}
The norms associated with these metrics are denoted
$\SPn{S/P}{\xi}=\sqrt{\SP{S/P}{\xi}{\xi}}$.

\begin{Proposition}
\label{P-is-rpt}
The scalar symbols
$q_{S/P}(t,x,\tau,\xi) =\rho(x)\big(\tau^2-\SP{S/P}{\xi}{\xi}\big)$
and their product $q_S q_P$ are of real principal type.
$P$ is a system of real principal type with principal symbol
\begin{equation}
\label{P-principal-symbol}
p= q_S (\Id_3-\pi) + q_P \pi.
\end{equation}
\end{Proposition}

\begin{Proof}
A straightforward computation gives
the principal symbol $p$ of $P$ at $(t,x,\tau,\xi)\in\Tstar(\spacetimeclosure)$
as follows:
\begin{equation*}
p=\rho\tau^2 \Id_3 -(\lambda+\mu)(\xi\otimes\xi) -\mu \xi^2\Id_3 -(\xi\cdot R\xi)\Id_3.
\end{equation*}
Hence $p$ has the asserted form.
It follows from \eqref{gP-less-gS} that $q_S$ and $q_P$
are of real principal type.
Furthermore
\begin{equation}
\label{qP-less-qS}
q_P(t,x,\tau,\xi) < q_S(t,x,\tau,\xi)
\quad\text{if $\xi\neq 0$.}
\end{equation}
Hence also $q=q_S q_P$ is a scalar symbol of real principal type.
Now $q^{-1}(0)={(\det p)}^{-1}(0)$ and
$\tilde{p}p=q\Id$ for $\tilde{p}= q_P (\Id_3-\pi) + q_S \pi$.
According to \cite[Definition~3.1]{dencker:82:polarization-sets}
$P$ is a system of real principal type
with characteristic variety $q=0$.
\end{Proof}

\begin{Remark}
Man~\cite{man:98:hartigs-law} proposes for elastodynamics
with residual stress $R$ a more general constituitive law
$S= R+\gradient u\cdot R+ C E$
where the elasticity tensor $C$ also depends linearly on $R$.
In the isotropic case $CE$ consists of the right-hand side in
\eqref{isotropic-C} plus the $R$ dependent terms
\begin{equation}
\label{isotropic-Ctilde}
\beta_1 \tr(E)\tr(R)I +\beta_2 \tr(R)E +\beta_3\big(\tr(E)R+\tr(ER)I\big)
	+\beta_4\big(ER+RE\big).
\end{equation}
In the inverse problem for real media the additional terms should
not be neglected since typically $R$ is much larger than the stress $S$.
A straightforward calculation shows that the elastodynamic
operator $P$ with this isotropic stress-strain relation is still of real principal type
in case $\beta_3=\beta_4=0$,
$\lambda+2\mu+\beta_1\tr(R) > \mu+\beta_2\tr(R)/2$,
and $(\mu(x)+\beta_2(x)\tr(R(x))/2){|\xi|}^2+R(x,\xi) >0$
when $(x,\xi)\in\Tstar(\closure{\Omega})\setminus 0$.
\end{Remark}

We recall some notions of the microlocal theory of
boundary problems and apply them to the system of elastodynamics.
Let $\gamma=(t,x,\tau,\xi_|)\in\Tstar(\spacetimebdry)\setminus 0$.
This means that there is given
$(t,x,\tau,\xi)\in\Tstar(\spacetimeclosure)$
with $x\in\boundary\Omega$ and
$\xi_| =\xi\restrictto{T_{x}(\boundary\Omega)}$.
$\gamma$ is called an elliptic, a hyperbolic, or a glancing point
of $S/P$ mode if the following quadratic equation in $z$,
$$
q_{S/P}(t,x,\tau,\xi-z\nu(x)) = 0,
$$
has no real roots, two distinct real roots, or a double real root,
respectively.
$\Tstar(\spacetimebdry)\setminus 0$ decomposes into the disjoint union
of the elliptic region $\Elliptic_{S/P}$,
the hyperbolic region $\Hyperbolic_{S/P}$,
and the glancing hypersurface $\Glancing_{S/P}$
of the $S/P$ mode.
Because of \eqref{qP-less-qS} we have
$\Elliptic_{S}\subset\Elliptic_{P}$ and
$\Hyperbolic_{P}\subset\Hyperbolic_{S}$.
$\Tstar(\spacetimebdry)\setminus 0$ is the disjoint union
of the hyperbolic region $\Hyperbolic_{P}$,
the mixed region $\Elliptic_{P}\cap\Hyperbolic_{S}$,
the elliptic region $\Elliptic_{S}$, and
the glancing set $\Glancing= \Glancing_S\cup \Glancing_P$.
The lens maps satisfy
$\ScattRel_{S/P}\subset \Hyperbolic_{S/P} \times \Hyperbolic_{S/P}$.

A simple real root $z$ is called forward (resp.\ backward) if the bicharacteristic
curve starting in $\xi-z\nu$ enters $\spacetime$ when time
increases (resp.\ decreases).
Characteristics and bicharacteristics are called forward or backward
correspondingly.
Observe from Hamiltons equations that a characteristic
$\xi-z\nu$, $z$ real, of $q_{S/P}$ is forward (resp.\ backward) if
$\tau \SP{S/P}{\xi-z\nu}{\nu}$ is positive (resp.\ negative).
We denote by $z_{S/P}= z_{S/P}(t,x,\tau,\xi,\nu)$
the forward real root $z$ or the complex root $z$ with
positive imaginary part of
$q_{S/P}(t,x,\tau,\xi-z\nu) = 0$.
We shall use the abbreviation $\xi_{S/P}=\xi-z_{S/P} \nu(x)$.

Given $\delta>0$ we define
\begin{equation}
\label{Gamma-delta}
\Gamma_\delta =\{(t,x,\tau,\xi_|)\in\Tstar(\spacetimebdry)\setminus 0\,;\,
    |\tau| \geq \delta |\xi_||\}.
\end{equation}
Here $|\xi_||=|\xi|$ if $\xi\cdot\nu=0$.
$\Lambda$ is pseudo-differential microlocally
at nonglancing points in $\Gamma_\delta$.

\begin{Proposition}
\label{P-microlocal-DN-op}
Let $L,\delta>0$.  Assume \eqref{Lame-bound}.
There exists $0<\varepsilon =\varepsilon(L,\delta)$
such that under the assumption \eqref{R-is-very-small}
the following holds.
Given $\gamma\in\Gamma_\delta\setminus\Glancing$
the DN map $\Lambda$ equals, in a microlocal neighbourhood
of $(\gamma,\gamma)$, a first order \pdo{} with
principal symbol given as follows.
\begin{equation}
\label{DN-principal-symbol}
\sigma(\Lambda): a\mapsto \lambda (a\cdot\xi') \nu +\mu(a\cdot\nu) \xi'
    +\mu(\xi'\cdot\nu) a + (R\xi'\cdot\nu)a
\end{equation}
when $a\in\CC\xi'$, resp.\ $a\cdot\xi'=0$,
with $\xi'=\xi_P$, resp.\ $\xi'=\xi_S$.
\end{Proposition}

We the need the following fact about the characteristics of $P$.

\begin{Lemma}
\label{P-Lopatinski}
Let $L,\delta>0$.  Assume \eqref{Lame-bound}.
There exists $0<\varepsilon =\varepsilon(L,\delta)$
such that the following is true if
\eqref{R-is-very-small} holds.
For $\gamma=(t,x,\tau,\xi_|)\in\Gamma_{\delta}\setminus\Glancing$
we have $z_S\neq z_P$,
${\xi_S}^2\neq 0$, ${\xi_P}^2\neq 0$, and
\begin{equation}
\label{non-zero-product}
\xi_S\cdot\xi_P \neq 0.
\end{equation}
\end{Lemma}

We prove Lemma~\ref{P-Lopatinski} in section~\ref{section:lemma}.

\begin{Proof}[of Proposition~\ref{P-microlocal-DN-op}]
Choose $\varepsilon$ as in Lemma~\ref{P-Lopatinski}.
Let $\gamma=(t,x,\tau,\xi_|)\in\Gamma_{\delta}\setminus\Glancing$.
Flatten the boundary $\boundary\Omega$ near $x$ with
a change of coordinates such that the differential at $x$ is orthogonal.

We consider the symbol $p=p(t,x,\cdot)$.
Its inverse is
$p^{-1}= q_S^{-1} (\Id_3-\pi) + q_P^{-1} \pi$.
Let $\gamma^+$ be a closed Jordan curve enclosing $z_S$ and $z_P$
but no other roots of $q(\tau,\xi-z\nu)=0$.
Observe $c_{S/P}:=(d/dz) q_{S/P}(\tau,\xi-z\nu) \restrictto{z=z_{S/P}} \neq 0$.
By the residue theorem
\begin{equation}
\label{P-residue-integral}
\begin{split}
A_j :&= \int_{\gamma^+} z^j\,p(\tau,\xi-z\nu)^{-1}\,\frac{dz}{2\pi i} \\
   &= (z_S^j/c_S)\, (\Id_3 -\pi(\xi_S)) + (z_P^j/c_P)\, \pi(\xi_P)
\end{split}
\end{equation}
for every non-negative integer $j$.
We show that $A_0$ is non-singular.
Assume $A_0 w=0$.
Then $0=\xi_S\cdot A_0 w= (\xi_S\cdot\xi_p)(\xi_P\cdot w)/(c_P \xi_P^2)$.
Applying Lemma~\ref{P-Lopatinski} we infer $\xi_P\cdot w=0$.
Therefore $w=\pi(\xi_S)w$ and $0=(\xi_P\cdot \xi_S)(\xi_S\cdot w)$.
Applying Lemma~\ref{P-Lopatinski} again we get $\xi_S\cdot w=0$.
Hence $w=\pi(\xi_S)w=0$.

The invertibility of $A_0$ implies that
\eqref{Lopatinski-for-Dirichlet-p} holds with $K=3$.
Inequality \eqref{Lopatinski-for-Dirichlet-g} holds because the rank
is bounded by $N=6$ minus the dimension of the eigenspaces
corresponding to eigenvalues not enclosed by $\gamma^+$
which is $3$.

We can now apply Proposition~\ref{singularities-in-Dirichlet}.
We find, microlocally near $\gamma$, a parametrix $U$
for the initial boundary problem.
Hence $\gamma\notin\WF\big(\Lambda f- (\nu\cdot S(Uf))\restrictto{\spacetimebdry}\big)$
if $\WF f$ is contained in a small conic neighbourhood of $\gamma$.
Here $S(u)$ denotes the stress tensor which corresponds
to the displacement $u$.
The displacement-to-traction map
$u\mapsto \nu\cdot S(u)$ is a first order differential operator
with principal symbol
\begin{equation}
\label{P-symbol-of-traction}
s(x,\xi):=
\lambda (\nu\otimes\xi) + \mu(\xi\otimes\nu)
    +\mu(\xi\cdot\nu)\Id_3 + (R\xi\cdot\nu)\Id_3.
\end{equation}
It then follows from 
Proposition~\ref{singularities-in-Dirichlet}
that $\Lambda\in\PDO{1}$ in a conic neigbourhood of $\gamma$.

It remains to prove the
formula~\eqref{DN-principal-symbol} for the principal symbol of $\Lambda$.
From \eqref{P-residue-integral} we obtain
\begin{equation}
\label{A-one}
A_1 v =
\begin{cases}
z_S A_0 v\in \xi_S^\bot & \text{if $v\in\xi_P^\bot$,} \\
z_P A_0 v\in \CC\xi_P & \text{if $v\in\CC \xi_S$.}
\end{cases}
\end{equation}
From~\eqref{symbol-of-DxU} we get a formula for the principal symbol
of the normal derivative $-D_\nu U$ followed by restriction to $\spacetimebdry$.
Using this, \eqref{P-symbol-of-traction}, and \eqref{A-one}
we deduce \eqref{DN-principal-symbol}.
\end{Proof}

\begin{Remark}
\label{remark-on-gamma-delta}
Given $L>0$ we choose $\delta>0$ such that
$\Tstar(\spacetimebdry)\setminus\Gamma_\delta\subset\Elliptic_S$
holds for every operator in $\LameClass(L,1/2)$.
We then choose $0<\varepsilon\leq 1/2$ such that for every
$P\in\LameClass(L,\varepsilon)$ the initial boundary problem with
Dirichlet boundary conditions is well-posed and
the assertions in Proposition~\ref{P-microlocal-DN-op} hold.
It follows from the proof of Proposition~\ref{P-microlocal-DN-op}
that microlocal forward and backward parametrices exist
in the hyperbolic, in the mixed, and in part of the elliptic region.
\end{Remark}

\section{Propagation of Polarization}
\label{section:polar}

Let $L>0$. Choose $0<\delta,\varepsilon$ as in
Remark~\ref{remark-on-gamma-delta}
at the end of section~\ref{section:elast}.
In the following we assume $P\in\LameClass(L,\varepsilon)$.

We analyze the polarization of solutions of $Pu=0$ by applying
polarization filters, i.e.,
certain approximate projection operators, to the Cauchy data of $u$.
Fix $E\in\PDO{1}(\spacetimebdry)$ scalar elliptic
with principal symbol $e$.
The Cauchy data of a solution of $Pu\in\Cinfty(\spacetimeclosure)$ are,
by definition,
$Cu = (Eu\restrictto{\spacetimebdry},\, \nu\cdot S\restrictto{\spacetimebdry})$.
Notice that the Cauchy data of a solution $u$ of the initial boundary value problem
$Pu=0$, $u=f$ at $\spacetimebdry$, and $u=0$ initially, are
represented using the DN map as follows:
$Cf=(Ef,\Lambda f)=Cu$.
Here, abusing notation, we also defined $Cf$.

We now describe, on the principal symbol level, the spaces onto which
polarization filters project.
Let $\gamma=(t,x;\tau,\xi_|)\in\Tstar(\spacetimebdry)\setminus 0$.
Set
$$
B_{S/P}^\pm(\gamma) =
\left(\begin{array}{c}
e(\gamma)\Id_3\\ s(x,\xi-z_{S/P}^\pm\nu)
\end{array}\right)
\ker p(t,x,\tau,\xi-z_{S/P}^\pm\nu)
\subset \CC^6
$$
if $\gamma\in\Hyperbolic_{S/P}$.
Here $z_{S/P}^{+}$ and $z_{S/P}^{-}$ are the forward and backward
roots of $q_{S/P}(t,x,\tau,\xi-z\nu) = 0$, respectively.
Also define the linear subspaces
$$
B_{S/P}(\gamma)= \sum
\left(\begin{array}{c}
e(\gamma)\Id_3\\ s(x,\xi-z\nu)
\end{array}\right)
\ker p(t,x,\tau,\xi-z\nu)
$$
where the sum ranges over the roots of
$q_{S/P}(t,x,\tau,\xi-z\nu)=0$.
Clearly, $B_{S/P}(\gamma)= B_{S/P}^+(\gamma) + B_{S/P}^-(\gamma)$
if $\gamma\in\Hyperbolic_{S/P}$.
The disjoint unions
$$
B_{S/P}^\pm = \dot{\cup}_{\gamma\in\Hyperbolic_{S/P}} B_{S/P}^\pm(\gamma)
\quad\text{and}\quad
B_{S/P} = \dot{\cup}_{\gamma\notin\Glancing_{S/P}} B_{S/P}(\gamma)
$$
are subsets of the trivial bundles, $\CC^6$.

\begin{Lemma}
\label{bundles-B}
$B_{S}^\pm$, resp.\ $B_{P}^\pm$, are vector subbundles of $\CC^6$ over
$\Hyperbolic_S$, resp.\ $\Hyperbolic_P$, of ranks $2$, resp.\ $1$.
$B_{P}$ is a vector subbundle of $\CC^6$ over 
$\Hyperbolic_P\cup\Elliptic_P$ of rank $2$.
Furthermore
\begin{equation}
\label{osum-B}
\begin{aligned}
\CC^6 &= B_S^+ \oplus B_S^- \oplus B_P^+ \oplus B_P^- 
	\quad\text{over $\Hyperbolic_P$,} \\
\CC^6 &= B_S^+ \oplus B_S^- \oplus B_P
	\quad\text{over $\Hyperbolic_S\cap\Elliptic_P$.}
\end{aligned}
\end{equation}
\end{Lemma}

\begin{Proof}
Given a point in $\boundary\Omega$
we introduce coordinates $x=(x_1,\ldots,x_n)=(x',x_n)$ such that $\Omega$ and
$\boundary\Omega$ correspond to $x_n>0$ and $x_n=0$, respectively.
Let $\xi=(\xi',\xi_n)$ denote the dual variables.
We also arrange that, at the given point in the coordinates,
formula~\eqref{P-symbol-of-traction} still holds.
At $x_n=0$,
\begin{equation*}
\left(\begin{array}{c}
e\Id_3 \\ s
\end{array}\right)
=
\left(\begin{array}{cc}
h_1\Id_3 & 0\\ * & \partial s/\partial \xi_n
\end{array}\right)
\left(\begin{array}{c}
(\tau^2+{\xi'}^2)^{1/2}\Id_3 \\ \xi_n\Id_3
\end{array}\right).
\end{equation*}
$\lambda, \mu\geq 0$ and \eqref{R-is-small}
imply the ellipticity of $\partial s/\partial\xi_n$.
Hence
\begin{equation}
\label{transform-Cauchy-data}
h(t,x',\tau,\xi')=
\left(\begin{array}{cc}
h_1\Id_3 & 0\\ * & \partial s/\partial \xi_n
\end{array}\right)
\end{equation}
defines an elliptic $6\times 6$ symbol of order $0$.
Therefore, it suffices to prove the Lemma when,
in the definitions of the vector spaces $B_{S/P}^{*}(\gamma)$,
the symbols $e$ and $s(x,\xi)$ are replaced by
$(\tau^2+{\xi'}^2)^{1/2}\Id_3$ and $\xi_n\Id_3$, respectively.
Having made this replacement the assertions follow
from spectral decomposition of the first order
symbol $g$ associated with $p$ in Lemma~\ref{reduction-to-first-order}
and from \eqref{map-kernels} together with the known dimensions of $\Kernel p$.
\end{Proof}

Let $\pi_{S/P}^\pm$ and $\pi_{S/P}$ denote the projectors
associated with the decompositions \eqref{osum-B}.
In the following $\Pi_{S/P}^\pm$ and $\Pi_{S/P}$
denote $6\times 6$ systems of \pdo s of order $0$
having principal symbols $\pi_{S/P}^\pm$ and $\pi_{S/P}$.
The operators are defined microlocally where the symbols are.

We now state how polarization in solutions can be tested
on the Cauchy data.

\begin{Proposition}
\label{polarization-analyzers}
Let $u\in{\Dprime(\spacetimeclosure)}^3$ such that
$Pu \in{\Cinfty(\spacetimeclosure)}^3$,
$Cu\in{\Hs{s-1}(\spacetimebdry)}^6$.
Let $\gamma\in\Tstar(\spacetimebdry)\setminus 0$.
If $\gamma\in\Hyperbolic_P$
then $\gamma\in\WFs{s}(\Pi_{P}^\pm Cu)$ if and only if
$\WFpols{s+1}(u)$ contains a Hamilton orbit above the
forward/backward compressional wave bicharacteristic which
issues from $\gamma$ into the interior.
If $\gamma\in\Hyperbolic_S\setminus\Glancing_P$
then $\gamma\in\WFs{s}(\Pi_{S}^\pm Cu)$ if and only if
$\WFpols{s+1}(u)$ contains a Hamilton orbit above the
forward/backward shear wave bicharacteristic which
issues from $\gamma$ into the interior.
\end{Proposition}

\begin{Proof}
Introduce coordinates $x=(x_1,\ldots,x_n)$ as in the
proof of Lemma~\ref{bundles-B}.
Abbreviate $D_x=(D',D_n)$.
The simplified Cauchy data
$$
C_0 u=(\Mod{D_t,D'}u\restrictto{x_n=0}, D_n u\restrictto{x_n=0})
$$
are related to the Cauchy data $Cu$ as follows:
$Cu\equiv HC_0 u$ where $H\in\PDO{0}(\spacetimebdry)^{6\times 6}$
with principal symbol equal to $h$ of equation~\eqref{transform-Cauchy-data}.
It now suffices to prove the Proposition with $Cu$ replaced by $C_0 u$
and the operators $\Pi_{S/P}^{\pm}$ replaced by $H^{-1}\Pi_{S/P}^{\pm}H$.
The assertions now follow from Proposition~\ref{WF-in-boundary-trace}
if we recall the argument in Remark~\ref{Remark-reduction-to-first-order}
of section~\ref{section:second}.
\end{Proof}

Curves which are bicharacteristics over the interior
and reflected at non-glancing boundary points,
with or without conversion between shear and compressional mode,
are called broken bicharateristics.
The propagation of singularities in the Cauchy data
is stated recursively as follows.

\begin{Proposition}
\label{propagation-in-cauchy-data}
Let $f\in\Hcs{s}(\spacetimebdry)$, $s\geq 3$, with
$\WFs{s+1}(f)\subset\Gamma_\delta$.
Let $T\in\RR$ such that no forward broken bicharacteristic
which issues from
$$
\WFs{s+1}(f)\cap \big(\WFs{s}(\Pi_S^+ Cf)\cup \WFs{s}(\Pi_P^+ Cf)\big)
$$
intersects $\Glancing\cap\{t\leq T\}$.
Then, after intersection with $\{t\leq T\}$,
\begin{equation}
\label{propagation-Cf}
\begin{split}
\WFs{s}(Cf)= \WFs{s+1}(f) &\cup
\ScattRel_S\big(\WFs{s}(\Pi_S^+ Cf)\big) \\
	&\cup \ScattRel_P\big(\WFs{s}(\Pi_P^+ Cf)\big).
\end{split}
\end{equation}
\end{Proposition}

\begin{Proof}
Let $u$ denote the solution of $Pu=0$ with Dirichlet
boundary value $f$ and zero initial data.
By \cite[Theorem~4.2]{dencker:82:polarization-sets}
and \cite{gerard:85:polarisation} the polarization set
$\WFpols{s+1}(u)\cap\{t\leq T\}$ is contained in the union
of Hamilton orbits which lie above the broken bicharacteristics
which issue from $\WFs{s+1}(f)$.
We apply Proposition~\ref{polarization-analyzers} at both
endpoints of bicharacteristics which connect boundary points.
We obtain
$\ScattRel_{S/P}\big(\WFs{s}(\Pi_{S/P}^+ Cf)\big)
  \subset\WFs{s}(\Pi_{S/P}^- Cf)$.
Therefore the right hand side of \eqref{propagation-Cf}
is contained in the left hand side.
To prove the opposite inclusion
let $\gamma\in\WFs{s}(Cf)\setminus \WFs{s+1}(f)$.
Then $\gamma$ is the endpoint of a forward bicharacteristic
contained in $\WFs{s+1}(u)$ and,
by Proposition~\ref{polarization-analyzers},
issued from $\WFs{s}(\Pi_S^+ Cf)$ or $\WFs{s}(\Pi_P^+ Cf)$.
\end{Proof}

Proposition~\ref{polarization-analyzers} combined with
the following result permits us to specify, without having to know
the coefficients of $P$, sources for
which compressional singularities are muted.

\begin{Proposition}
\label{mute-P-polarization}
Choose $M\in\PDO{0}(\spacetimebdry)^{3\times 3}$ such that its
principal symbol $m$ equals at every
$(t,x,\tau,\xi_|)\in\Tstar(\spacetimebdry)$
the orthogonal projector onto the
one-dimensional subspace of $\RR^3$ which is orthogonal to $\xi$
and $\nu(x)$.
Let $\gamma\in \Hyperbolic_P$.
There is a conic neighbourhood $\Gamma\subset \Hyperbolic_P$
of $\gamma$ such that the following inclusion holds for every
$f\in\Hcs{s}(\spacetimebdry)^3$, $s\geq 3$, with
$\WF(f)\subset \Gamma$:
\begin{equation}
\label{mute-WF}
\WFs{s}\big(\Pi_P^+ C f \big) \subset \WFs{s+1}(f-M f).
\end{equation}
\end{Proposition}

\begin{Proof}
First we show
\begin{equation}
\label{image-of-double-M}
\left(\begin{array}{cc} m& 0\\ 0& m \end{array}\right)
\CC^6
\subset B_S^+ + B_S^-.
\end{equation}
To see this let $\gamma=(t,x,\tau,\xi_|)$ and
$a\in\CC^3$ with $\xi\cdot a=\nu\cdot a =0$, $\nu=\nu(x)$.
In view of \eqref{P-principal-symbol}
$a$ belongs to the kernel of $p(t,x,\tau,\xi_S^\pm)$.
In view of the definition \eqref{P-symbol-of-traction}
we have $s(x,\xi_S^\pm)a= \SP{S}{\xi_S^\pm}{\nu} a$.
Hence $(e(\gamma) a, \SP{S}{\xi_S^\pm}{\nu} a) \in B_S^\pm(\gamma)$.
$\SP{S}{\xi_S^+ - \xi_S^-}{\nu}\neq 0$ because $z_S^+\neq z_S^-$.
Therefore we obtain $(0,a),(a,0) \in B_S^+(\gamma) + B_S^-(\gamma)$
proving \eqref{image-of-double-M}.

$\pi_{P}= \pi_{P}^+ + \pi_{P}^-$ vanishes on $B_S^+ + B_S^-$.
Therefore \eqref{image-of-double-M} and the symbol calculus imply
\begin{equation}
\label{projectors-and-M}
\Pi_P \left(\begin{array}{cc} M& 0\\ 0& M \end{array}\right)
\in \PDO{-1}.
\end{equation}

Choose the conic neighbourhood $\Gamma$ of $\gamma$
in such a way that the DN map $\Lambda$ is a \pdo{} in
$\Gamma\times\Gamma$.
Shrinking $\Gamma$ if necessary we may assume that
every solution of $Pu=0$ which has zero initial data
and Dirichlet data $f$ with $\WF(f)\subset\Gamma$
does not contain backward bicharacteristics issuing
from $\Gamma$ in its wavefront set.

Observe from formula~\eqref{DN-principal-symbol}
that the principal symbol of $\Lambda$ maps the space
onto which $m$ projects into itself.
Hence $\Lambda M -M\Lambda M\in\PDO{0}$ and therefore
\begin{equation}
\label{CM-formula}
\left(\begin{array}{c}
EM\\ \Lambda M
\end{array}\right)
\equiv
\left(\begin{array}{cc}
M& 0\\ 0& M
\end{array}\right)
\left(\begin{array}{c}
E\Id_3\\ \Lambda M
\end{array}\right)
\quad \mod \PDO{0}.
\end{equation}
Let $f\in\Hcs{s}(\spacetimebdry)^3$, $s\geq 3$, with
$\WF(f)\subset \Gamma$.
Equations \eqref{projectors-and-M} and \eqref{CM-formula} imply
\begin{equation*}
\Pi_P \left(\begin{array}{c} EMf\\ \Lambda Mf \end{array}\right)
\in \Hs{s}(\spacetimebdry).
\end{equation*}
Assume $\gamma\notin\WFs{s+1}(f-M f)$.
Then, recalling $Cf=(Ef, \Lambda f)$, we obtain
\begin{equation}
\label{Cu-Mf}
\gamma\notin\WFs{s}(\Pi_P Cf).
\end{equation}
Proposition~\ref{polarization-analyzers} and our choice of $\Gamma$
imply that \eqref{Cu-Mf} holds with $\Pi_P$ replaced by $\Pi_P^-$.
Hence \eqref{Cu-Mf} also holds with $\Pi_P$ replaced by $\Pi_P^+$.
\end{Proof}

\section{Proof of Theorem~\ref{thm:recover-relations}}
\label{section:proof}

The idea is to recover the lens maps from the elements
of $\WF(\Lambda f)\setminus\WF(f)$
with least time where $f$ ranges over point sources.

Let $L>0$. Choose $0<\delta,\varepsilon$ as in
Remark~\ref{remark-on-gamma-delta}
at the end of section~\ref{section:elast}.
Let $P^{(1)}, P^{(2)}\in\LameClass(L,\varepsilon)$.
Assume $\Lambda^{(1)} = \Lambda^{(2)}$.
We show that the shear and compressional lens maps are equal:
$\ScattRel_{S}^{(1)} = \ScattRel_{S}^{(2)}$ and
$\ScattRel_{P}^{(1)} = \ScattRel_{P}^{(2)}$.
Let $\tilde{\ScattRel}$ denote the union of the sets
$\ScattRel_{S/P}^{(j)} \cap \big(
(\Glancing^{(k)}\times\Gamma_\delta)\cup (\Gamma_\delta\times\Glancing^{(k)}) \big)$.
Observe that
$\ScattRel_{S/P}^{(j)}\setminus\tilde{\ScattRel}$
is open and dense in $\ScattRel_{S/P}^{(j)}$.
We first prove that the lens maps agree outside $\tilde{\ScattRel}$.

Fix $s\geq 3$.
Below we choose, given $\gamma^\In\in\Gamma_\delta$ nonglancing,
sources $f$ with the properties
\begin{equation}
\label{source-f}
f\in \Hcs{s}(\spacetimebdry)^3,\quad
\WFs{s+1}(f) = \RR_+ \gamma^{\In}.
\end{equation}
We study the singularities of the Cauchy data
$Cf=(Ef,\Lambda^{(1)} f) =(Ef,\Lambda^{(2)} f)$.
Because of the zero initial condition the backward bicharacteristics
issuing from $\gamma^\In$ are disjoint from $\WFs{s+1}(u)$.
By Proposition~\ref{polarization-analyzers} 
$\gamma^{\In}\notin \WFs{s}(\Pi_{S/P}^{-(j)} Cf)$.
$\gamma^\In\in\WFs{s}(Cf)$ by assumption
on $f$ since $E$ is elliptic.
Therefore
$\gamma^{\In}\in \WFs{s}(\Pi_{S}^{+(j)} Cf)
   \cup \WFs{s}(\Pi_{P}^{+(j)} Cf) \cup \Elliptic_P^{(j)}$.

Let $(\gamma^{\Out},\gamma^{\In})\in\ScattRel_{S}^{(1)}\setminus\tilde{\ScattRel}$.
Choose $f$ with \eqref{source-f} and
$\gamma^{\In}\notin \WFs{s+1}(f-Mf)$.
Proposition~\ref{mute-P-polarization} implies
$\gamma^{\In}\notin \WFs{s}(\Pi_P^{+(j)} Cf)$
for $j=1$ and $j=2$.
Consequently, $\gamma^{\In}\in \WFs{s}(\Pi_S^{+(j)} Cf)$.
Proposition~\ref{propagation-in-cauchy-data} with $P^{(1)}$ implies
\begin{equation}
\label{WFofCfequalsInandOut}
\WFs{s}(Cf)\cap\{t\leq t(\gamma^\Out)\} =
   \RR_+ \gamma^\In \cup \RR_+ \gamma^\Out.
\end{equation}
\eqref{WFofCfequalsInandOut} and
Proposition~\ref{propagation-in-cauchy-data} with $P^{(2)}$ imply
$(a\gamma^{\Out},\gamma^{\In})\in\ScattRel_{S}^{(2)}$ for some $a>0$.
The covariable $\tau$ is constant along bicharacteristics.
Therefore $a=1$.
Thus we have shown
$\ScattRel_{S}^{(1)}\setminus \tilde{\ScattRel} \subset \ScattRel_{S}^{(2)}$.
Interchanging $P^{(1)}$ with $P^{(2)}$ we obtain
\begin{equation}
\label{S-lenses-almost-equal}
\ScattRel_{S}^{(1)}\setminus \tilde{\ScattRel} =
   \ScattRel_{S}^{(2)}\setminus \tilde{\ScattRel}.
\end{equation}

Let $(\gamma^{\Out},\gamma^{\In})\in\ScattRel_{P}^{(1)}\setminus\tilde{\ScattRel}$.
Choose $f$ with \eqref{source-f} and
\begin{equation}
\label{WfcompressionalOne}
\gamma^{\In}\in \WFs{s}(\Pi_P^{+(1)} Cf)\setminus \WFs{s}(\Pi_S^{+(1)} Cf).
\end{equation}
Proposition~\ref{propagation-in-cauchy-data} with $P^{(1)}$ implies
\eqref{WFofCfequalsInandOut}.
\eqref{WFofCfequalsInandOut} and
Proposition~\ref{propagation-in-cauchy-data} with $P^{(2)}$ imply
$(a\gamma^{\Out},\gamma^{\In})\in \ScattRel_{S}^{(2)}\cup \ScattRel_{P}^{(2)}$
with $a>0$. Again $a=1$ follows.
If $(\gamma^{\Out},\gamma^{\In})\in \ScattRel_{S}^{(2)}$
then also
$(\gamma^{\Out},\gamma^{\In})\in \ScattRel_{S}^{(1)}$
by \eqref{S-lenses-almost-equal}.
Hence $(\gamma^{\Out},\gamma^{\In})\in \ScattRel_{P}^{(2)}$ if
$(\gamma^{\Out},\gamma^{\In})\notin \ScattRel_{S}^{(1)}$.
Suppose $(\gamma^{\Out},\gamma^{\In})\in \ScattRel_{S}^{(1)}$.
Choose $f$ as above but now with projectors $\Pi_{S/P}^{+(1)}$
in \eqref{WfcompressionalOne} replaced by $\Pi_{S/P}^{+(2)}$.
Equation~\eqref{WFofCfequalsInandOut} still follows from
Proposition~\ref{propagation-in-cauchy-data} with $P^{(1)}$ because
$(\gamma^{\Out},\gamma^{\In})\in \ScattRel_{S}^{(1)}\cap \ScattRel_{P}^{(1)}$.
$\gamma^{\In}\notin \WFs{s}(\Pi_S^{+(2)} Cf)$,
\eqref{WFofCfequalsInandOut}, and
Proposition~\ref{propagation-in-cauchy-data} with $P^{(2)}$ imply
$(\gamma^{\Out},\gamma^{\In})\in \ScattRel_{P}^{(2)}$.
Thus we have shown
$\ScattRel_{P}^{(1)}\setminus \tilde{\ScattRel} \subset \ScattRel_{P}^{(2)}$.
Interchanging $P^{(1)}$ with $P^{(2)}$ we obtain
\begin{equation}
\label{P-lenses-almost-equal}
\ScattRel_{P}^{(1)}\setminus \tilde{\ScattRel} =
   \ScattRel_{P}^{(2)}\setminus \tilde{\ScattRel}.
\end{equation}

It remains to show that
\eqref{S-lenses-almost-equal} and \eqref{P-lenses-almost-equal}
hold with $\tilde{\ScattRel}$ replaced by the empty set.
Assume $(\gamma^{\Out},\gamma^{\In})\in\ScattRel_{S}^{(1)}$.
We use a limit argument to prove
$(\gamma^{\Out},\gamma^{\In})\in\ScattRel_{S}^{(2)}$.
First we observe that
$\gamma^\In, \gamma^\Out\notin\Glancing_S^{(2)}$.
Suppose not. 
Then a neighbourhood of
$(\gamma^{\Out},\gamma^{\In})$ in $\ScattRel_{S}^{(1)}$
contains a point which is in
$(\Gamma_\delta\times\Elliptic_S^{(2)})\setminus\tilde{\ScattRel}$
or in
$(\Elliptic_S^{(2)}\times\Gamma_\delta)\setminus\tilde{\ScattRel}$.
This point cannot be in $\ScattRel_{S}^{(2)}$.
This contradicts \eqref{S-lenses-almost-equal}.
Choose a sequence
$(\gamma^{\Out}_k,\gamma^{\In}_k)\in \ScattRel_{S}^{(1)}\cap \ScattRel_{S}^{(2)}$,
$k\in\NN$, which converges to $(\gamma^{\Out},\gamma^{\In})$.
Let $\gamma_k$ denote the shear wave bicharacteristic
for $P^{(2)}$  with $\gamma_k(0)=\gamma^{\In}_k$ and
$\gamma_k(t_k)=\gamma^{\Out}_k$, $t_k>0$.
The length sequence of the corresponding sequence of geodesics is bounded.
By compactness there is a limit geodesic and thus a bicharacteristic
$\gamma:[0,T]\rightarrow\Tstar(\spacetimeclosure)$ of $P^{(2)}$
with $\gamma(0)=\gamma^\In$ and $\gamma(T)=\gamma^\Out$.
It suffices to show $\gamma(t)$ lies over the interior when $0<t<T$.
Suppose we had $\gamma(t^*)$ above the boundary for some $0<t^*<T$.
Then every neighbourhood of $(\gamma(t^*),\gamma^\In)$
has non-empty intersection with $\ScattRel_{S}^{(2)}$ hence,
by \eqref{S-lenses-almost-equal}, also with $\ScattRel_{S}^{(1)}$.
This contradicts the continuity of the map $\ScattRel_{S}^{(1)}$
at $\gamma^\In$.
Hence we have shown $\ScattRel_{S}^{(1)}\subset \ScattRel_{S}^{(2)}$.
The other inclusions are proved in the same way.

\section{Proof of Lemma~\ref{P-Lopatinski}}
\label{section:lemma}

Let $\gamma=(t,x,\tau,\xi_|)\in\Gamma_{\delta}\setminus\Glancing$.
To ease notation we drop the coordinates $(t,x)$.

Recall the definitions of the symbols
$q_{S/P}(\tau,\xi)/\rho=\tau^2-|\xi|_{S/P}^2$,
the metrics
$\rho|\xi|_{P}^2= \rho|\xi|_{S}^2+(\lambda+\mu)|\xi|^2$,
$\rho|\xi|_{S}^2=\mu|\xi|^2+ R\xi\cdot\xi$,
and the characteristics $\xi_{S/P}=\xi-z_{S/P}\nu$,
$q_{S/P}(\tau,\xi_{S/P})=0$.
We have $|\nu|=1$.

The equation
$q_S(\tau,\xi-z\nu) - q_P(\tau,\xi-z\nu) = (\lambda+\mu) (\xi-z\nu)^2$
holds for $z\in\CC$.
It implies $\xi_S\cdot\xi_P=\xi_{S/P}^2 =0$ if $z_S=z_P$.
Also it implies $q_P(\tau,\xi_S)=0$ if $\xi_{S}^2 = 0$.
So if we had $\xi_{S}^2 = 0$ then $z_S\notin\RR$ because
of the real principal type property of $q_S q_P$.
Since $\IM z_{S/P}\geq 0$ this can only happen if $z_P=z_S$.
Therefore $\xi_{S}^2 = 0$ implies $\xi_S\cdot\xi_P=0$.
In the same way $\xi_{P}^2 = 0$ implies $\xi_S\cdot\xi_P=0$.
Therefore, it suffices to prove the inequality \eqref{non-zero-product}.

Choose $0<\varepsilon \leq 1/2$.
$\varepsilon>0$ will be decreased further depending on $L$ and $\delta$ only.
The smallness assumption \eqref{R-is-very-small} 
on the residual stress tensor implies
\begin{equation}
\label{eps-less-than-one}
1-\varepsilon\leq \frac{\rho|\eta|_S^2}{\mu|\eta|^2},\;
 \frac{\rho|\eta|_P^2}{(\lambda+2\mu)|\eta|^2}\leq 1+\varepsilon
\quad\text{if $\eta\neq 0$.}
\end{equation}

Consider the elliptic case,
$\gamma\in\Elliptic_S\subset\Elliptic_P$.
Without loss of generality we assume $\xi\cdot\nu=0$.
Then $\xi_S\cdot\xi_P=|\xi|^2+z_S z_P$ and $|\xi|=|\xi_||\leq|\tau|/\delta$.
$(\lambda+2\mu)/\rho\leq L^2$ by \eqref{Lame-bound}.
Therefore, we assume
\begin{equation}
\label{eps-tau}
\rho\tau^2> 2 \varepsilon (\lambda+2\mu)|\xi|^2
 \geq 2 \varepsilon \mu|\xi|^2.
\end{equation}
$z=z_{S/P}$ is the solution with positive imaginary part
of the quadratic equation
$$
|\nu|_{S/P}^2 z^2 -2b z +(|\xi|_{S/P}^2 -\tau^2) = 0
\quad\text{where $b=R\xi\cdot\nu$.}
$$
Notice that the signs of the real parts of $z_S$ and $z_P$ are equal to the sign of $b$.
Hence $z_S z_P\notin\RR$ and thus $\xi_S\cdot\xi_P\neq 0$ if $b\neq 0$.
Assume $b=0$.
We solve the quadratic equations and then estimate using
\eqref{eps-less-than-one} and \eqref{eps-tau}:
\begin{align*}
|z_S z_P|^2
& = \frac{\rho(|\xi|_S^2-\tau^2)}{\rho|\nu|_S^2}\cdot
    \frac{\rho(|\xi|_P^2-\tau^2)}{\rho|\nu|_P^2}
\\
 & <
  \frac{\big(\mu(1+\varepsilon) - 2\varepsilon\mu\big)|\xi|^2}{(1-\varepsilon)\mu|\nu|^2}
  \cdot
  \frac{\big((\lambda+2\mu)(1+\varepsilon) - 2\varepsilon(\lambda+2\mu)\big)|\xi|^2}%
       {(1-\varepsilon)(\lambda+2\mu)|\nu|^2}
\\
 & \leq |\xi|^4.
\end{align*}
Hence $|\xi|^2+z_S z_P\neq 0$, i.e., \eqref{non-zero-product} holds.

In the mixed case, $\gamma\in\Elliptic_P\cap\Hyperbolic_S$,
we have $z_S\in\RR$ and $\IM z_P>0$.
This implies \eqref{non-zero-product}.

Consider the hyperbolic case,
$\gamma\in\Hyperbolic_P\subset\Hyperbolic_S$.
Without loss of generality we assume
\begin{equation}
\label{nu-xi-P-is-zero}
\SP{P}{\xi}{\nu}=0.
\end{equation}
Then the roots of
$0= \tau^2 -|\xi-z\nu|_P^2 = -\SPn{P}{\nu}^2 z^2 -\SPn{P}{\xi}^2 +\tau^2$
have opposite signs.
Since $q_P<q_S$ this is also true for the roots
of $0= \tau^2-|\xi-z\nu|_S^2$.
Furthermore $0<|z_P|<|z_S|$.
Since $z_S$ and $z_P$ are both forward they have the same sign.
For simplicity we assume $0<z_P< z_S$.
We shall use the estimate
\begin{equation}
\label{hyp-xip-xis-estimate}
\xi_S\cdot \xi_P \geq |\xi|^2 - \varepsilon z_S |\xi| +z_S z_P.
\end{equation}
$(\lambda+2\mu)\xi\cdot\nu +R\xi\cdot\nu=0$ is
equation~\eqref{nu-xi-P-is-zero} restated.
From this and \eqref{R-is-very-small} we deduce
$2|\xi\cdot\nu|\leq \varepsilon |\xi|$.
Hence 
$|(z_S+z_P)(\xi\cdot\nu)|\leq 2z_S|\xi\cdot\nu| \leq \varepsilon z_S |\xi|$.
\eqref{hyp-xip-xis-estimate} follows.

The equation
$q_{S}(\tau,\xi-z_S\nu) -q_{P}(\tau,\xi-z_P\nu)=0$
is equivalent to
\begin{equation}
\label{quadratic-equn-for-zS}
z_S^2 ( |\nu|_S^2 -t^2 |\nu|_P^2)
-2 z_S \SP{S}{\xi}{\nu} -((\lambda+\mu)/\rho)|\xi|^2=0
\quad\text{where $t=z_P/z_S$.}
\end{equation}
We estimate the root $z_S$ of this quadratic equation.
Using the Cauchy-Schwarz inequality and \eqref{eps-less-than-one}
to estimate $|\nu|_S$ from below we deduce
from~\eqref{quadratic-equn-for-zS}
$$
|z_S|/4 \leq |\xi|_S + |\xi|\sqrt{(\lambda+\mu)/ \mu} 
\quad\text{if $2t\SPn{P}{\nu}\leq \SPn{S}{\nu}$.}
$$
Decreasing $\varepsilon>0$ if necessary, we assume
$\varepsilon z_S \leq |\xi|$ if $2t\SPn{P}{\nu}\leq \SPn{S}{\nu}$.
Inserting this estimate into \eqref{hyp-xip-xis-estimate}
we get $\xi_S\cdot\xi_P\geq z_S z_P>0$.
It remains to prove \eqref{non-zero-product}
when $2t\SPn{P}{\nu} > \SPn{S}{\nu}$.
From \eqref{eps-less-than-one} and \eqref{Lame-bound} we get
$\SPn{P}{\nu}/\SPn{S}{\nu}\leq 3(\lambda+2\mu)/\mu\leq 3L^2$.
Decreasing $\varepsilon>0$ if necessary, we assume
$\varepsilon^2 \SPn{P}{\nu} \leq 2 \SPn{S}{\nu}$.
Hence $\varepsilon^2 z_S^2/4<z_S z_P$.
We estimate the right hand side of \eqref{hyp-xip-xis-estimate}
from below and get
$\xi_S\cdot \xi_P  > (|\xi| - \varepsilon z_S/2)^2 \geq 0$.
\bibliographystyle{amsalpha}

\providecommand{\bysame}{\leavevmode\hbox to3em{\hrulefill}\thinspace}

\end{document}